\documentclass[letter]{aa}
\usepackage{amssymb,amsmath}
\usepackage{graphicx}
\usepackage{xcolor,natbib}
\usepackage{multirow}
\usepackage[T1]{fontenc}
\usepackage{ae,aecompl}
\usepackage{newtxtext, newtxmath}
\usepackage{float}
\usepackage{subfigure}
\usepackage{longtable}
\usepackage{enumitem}
\usepackage{longtable,listings}
\usepackage[flushleft]{threeparttable}
\usepackage{parcolumns}
\usepackage{hyperref}

\def\oc3{[O~{\sc iii}]$_c$}
\def\ob3{[O~{\sc iii}]$_b$}

\begin{document}

\title{Periodic variations of the first and second moments of broad Balmer emission lines from central accretion disks}

\titlerunning{accretion disk origin of broad emission lines}


\author{XueGuang Zhang}

\institute{Guangxi Key Laboratory for Relativistic Astrophysics, School of Physical Science and Technology, 
GuangXi University, No. 100, Daxue Road, Nanning, 530004, P. R. China \ \ \ \email{xgzhang@gxu.edu.cn}}

\abstract
{Broad emission line regions (BLRs) lying into central accretion disks has been widely accepted to explain the unique double-peaked 
broad emission lines in Active Galactic Nuclei (double-peaked BLAGNs). Here, accepted the accretion disk origin, periodic variations 
of central wavelength $\lambda_{0}$ (the first moment) and line width $\sigma$ (the second moment) of double-peaked broad emission 
lines are theoretically simulated and determined. Furthermore, through theoretically simulated periodicities of $T_{\lambda0}$ and 
$T_{\sigma}$ for variations of $\lambda_0$ and $\sigma$, periodicity ratio $R_{fs}$ of $T_{\lambda0}$ to $T_{\sigma}$ to be around 
2 can be applied to support the spiral arms to be more preferred in BLRs lying into central accretion disks. Then, periodic 
variations of $\lambda_0$ and $\sigma$ are determined and shown in the known double-peaked BLAGN NGC1097, leading to the parameter 
$R_{fs}\sim2$, which can be applied as clues to support that the structure of spiral arms in disk-like BLRs in central accretion 
disk should be the most compelling interpretation to the variability of double-peaked broad H$\alpha$ in NGC1097. The results provide 
clean criteria to test accretion disk origins of double-peaked broad emission lines in AGN.}


\maketitle

\section{Introduction}

	Among the broad emission line Active Galactic Nuclei (BLAGNs), there is one special subclass, the BLAGNs with double-peaked 
broad emission lines (double-peaked BLAGNs). The first reports on the double-peaked BLAGNs can be found in the 1980s, such as in 
\citet{ch89a, ch89b}. Since 1980s, there are large samples of double-peaked BLAGNs, such as the 12 double-peaked BLAGNs in \citet{eh94}, 
the 116 double-peaked BLAGNs in \citet{st03}, the more recent 250 double-peaked BLAGNs in \citet{wg24}. In order to explain the 
double-peaked broad emission lines, two main models have been proposed, the model related to binary black hole (BBH) system (the 
BBH model) and the model related to broad emission line regions (BLRs) lying into central accretion disk (the accretion disk model).

	The BBH model has been firstly proposed to explain the systematic variability of the double-peaked broad Balmer lines in 
3C390.3 in \citet{ga96}. However, in \citet{eh97} through spectroscopic monitoring of 3C 390.3 spanning two decades, the BBH model 
has been ruled out, due to no long-term systematic changes in radial velocity expected by the BBH model. Besides the BBH model, 
the accretion disk model has been accepted to the double-peaked broad emission lines, such as the circular accretion disk model 
in \citet{ch89a, ch89b}, the elliptical accretion disk model in \citet{el95}, the accretion disk model considering warped structures 
in \citet{hb00}, the circular accretion disk plus arms model in \citet{se03}, the stochastically perturbed accretion disk model 
in \citet{fe08}, etc.

\begin{figure*}
\centering\includegraphics[width = 18cm,height=7cm]{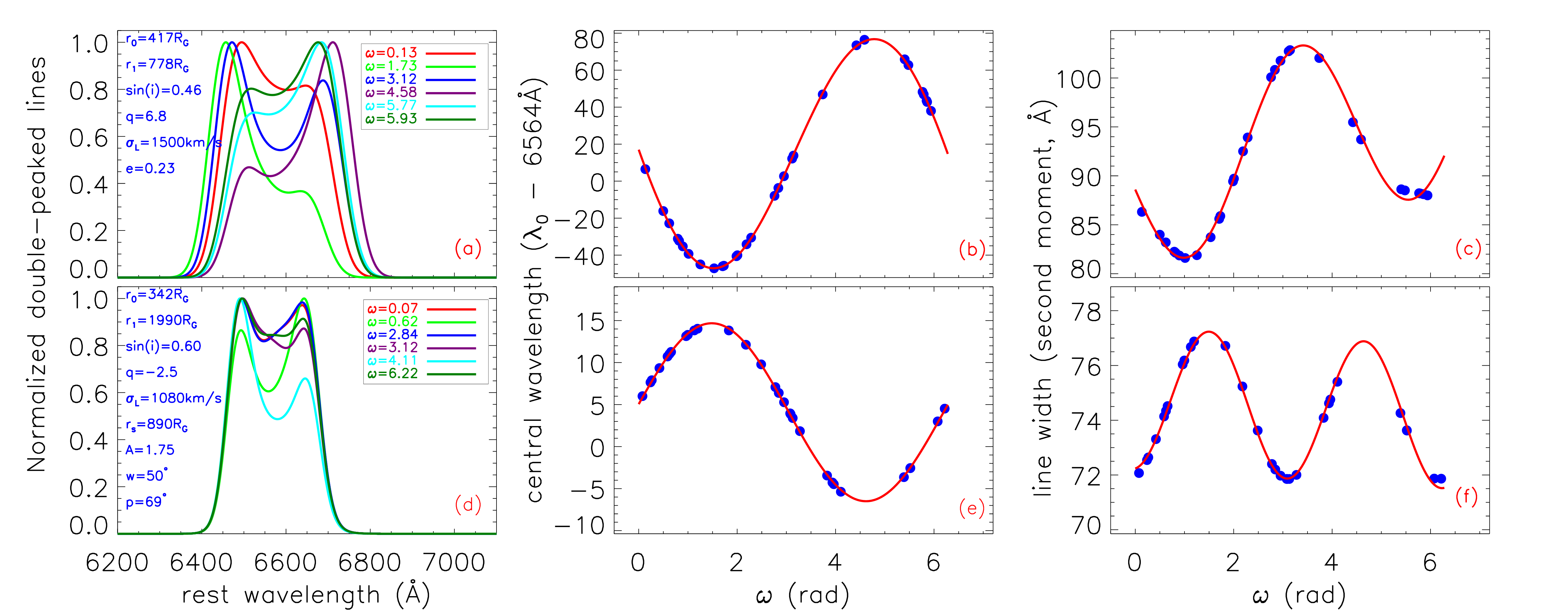}
\caption{Panel (a) shows examples of the simulated double-peaked broad emission lines in different phases through the elliptical 
accretion disk model. The solid lines in different colors show the line profiles in different phases with different orientation 
angles $\omega$, as shown in the legend in top right corner. Panel (b) and (c) show the dependence of the first moment $\lambda_{0}$ 
and the second moment $\sigma$ on the orientation angle $\omega$ for the case shown in panel (a). In panel (b) and (c), solid blue 
circles show the $\lambda_{0}$ and $\sigma$ calculated through the simulated results with the randomly collected 30 values of 
$\omega$, solid red line shows the best descriptions to the $\lambda_{0}(\omega)$ and $\sigma(\omega)$ by applications of a sine 
function plus a linear trend. Panel (d), (e) and (f) show the corresponding results through the circular accretion disk plus arms 
model.
}
\label{model}
\end{figure*}

	Since the accretion disk models proposed, more and more studies have shown that the current models are not enough but the 
elliptical accretion disk model and/or the circular accretion disk plus arms model should be the preferred one to explain the 
emission features of the double-peaked broad lines, especially through the long-term variability of the double-peaked broad lines, 
such as the discussions in \citet{se03, el09, ge07, le10, ss12, zh13}. And the circular accretion disk plus arms model and/or 
the elliptical accretion disk model have been well applied to describe the double-peaked broad lines in the literature, as discussed 
and shown in \citet{ss17, hf20, zh21, zh22, wg24, zh24a, zh24}.

	Meanwhile, when studying the line profile variability of the double-peaked broad emission lines, the following parameters 
have been well checked, the positions and intensities of the two peaks, the line widths (full width of the profile at half-maximum 
and/or quarter-maximum) of the double-peaked broad lines, etc., and quasi-periodic variations of the line parameters could be 
basically expected by the accretion disk model, such as the simple results shown in Fig.31-36 in \citet{le10}. However, the 
expected quasi-periodic variations are not well consistent with results from the multi-epoch spectroscopic results, probably due 
to the following two main reasons. On the one hand, the peak positions are not apparent in the observational line spectra, leading 
to large uncertainties of the peak positions and/or peak intensities. On the other hand, probably the peak of the double-peaked 
broad lines mixed with narrow emission lines lead to apparent effects of the determined line widths.

	In this manuscript, not the information of peaks of the double-peaked broad emission lines, but the first moment (central 
wavelength, $\lambda_0$) and the second moment (line width, $\sigma$) of the double-peaked broad emission lines are checked through 
the accretion disk models and then compared with the observational results. There are few effects of the peak positions of the line 
profiles on estimations of $\lambda_0$ and $\sigma$, probably leading to more apparent periodic variations of $\lambda_0$ and 
$\sigma$, which is our main objective. Section 2 presents our main results and necessary discussions on variability of $\lambda_0$ 
and $\sigma$ of double-peaked broad lines. Section 3 gives our main conclusions. And in this manuscript, we have adopted the 
cosmological parameters of $H_{0}=70{\rm km\cdot s}^{-1}{\rm Mpc}^{-1}$, $\Omega_{\Lambda}=0.7$ and $\Omega_{\rm m}=0.3$.

\section{Main Results and Necessary Discussions} 

\begin{figure*}
\centering\includegraphics[width = 18cm,height=6cm]{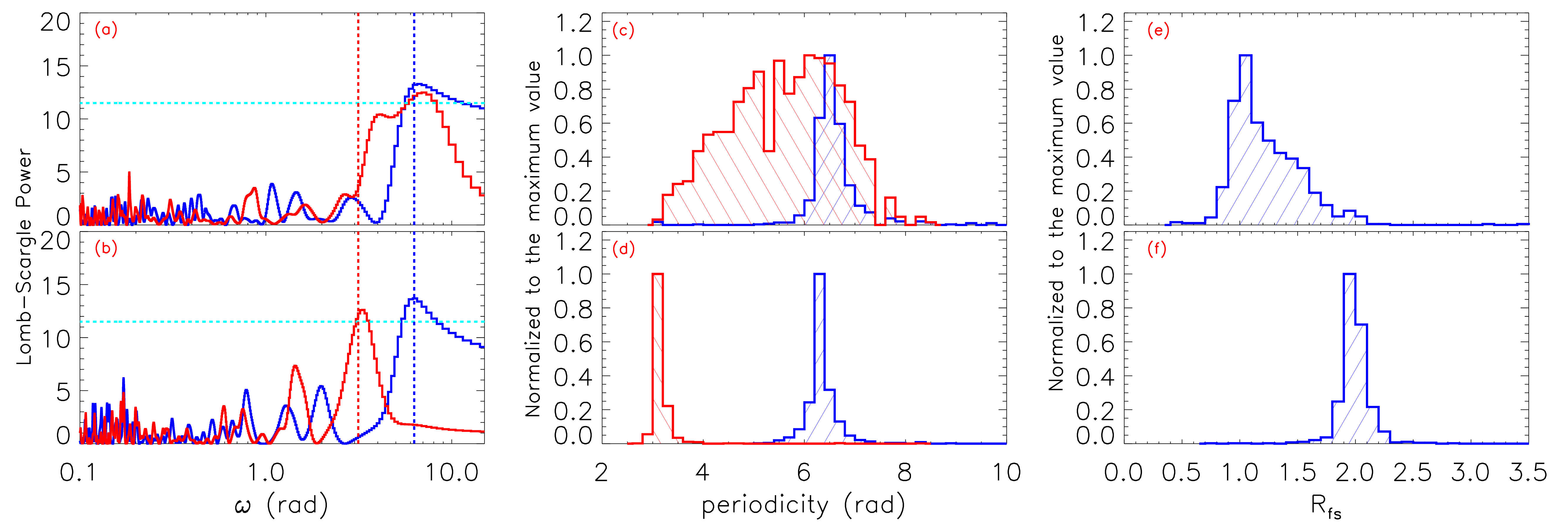}
\caption{Panel (a), (b) show the Lomb-Scargle power of the $\lambda_{0}(\omega)$ (solid blue line) and $\sigma(\omega)$ (solid red 
line) shown in panel (b), (c) and (e), (f) in Fig.~\ref{model}. Horizontal cyan dashed lines mark the confidence level of 99\% 
(corresponding false alarm probability 0.01) for periodicities. Panel (c), (d) show the periodicity (in units of rad) distributions 
for the 864, 894 cases through the elliptical accretion disk model and the circular accretion disk plus arms model, respectively. 
Histograms in blue and in red show the results for the periodicity distributions in $\lambda_0(\omega)$ and in $\sigma(\omega)$, 
respectively. Panel (e), (f) show the distributions of $R_{fs}$ through the elliptical accretion disk model and the circular 
accretion disk plus arms model, respectively.
}
\label{gls}
\end{figure*}

\begin{figure*}
\centering\includegraphics[width = 18cm,height=8cm]{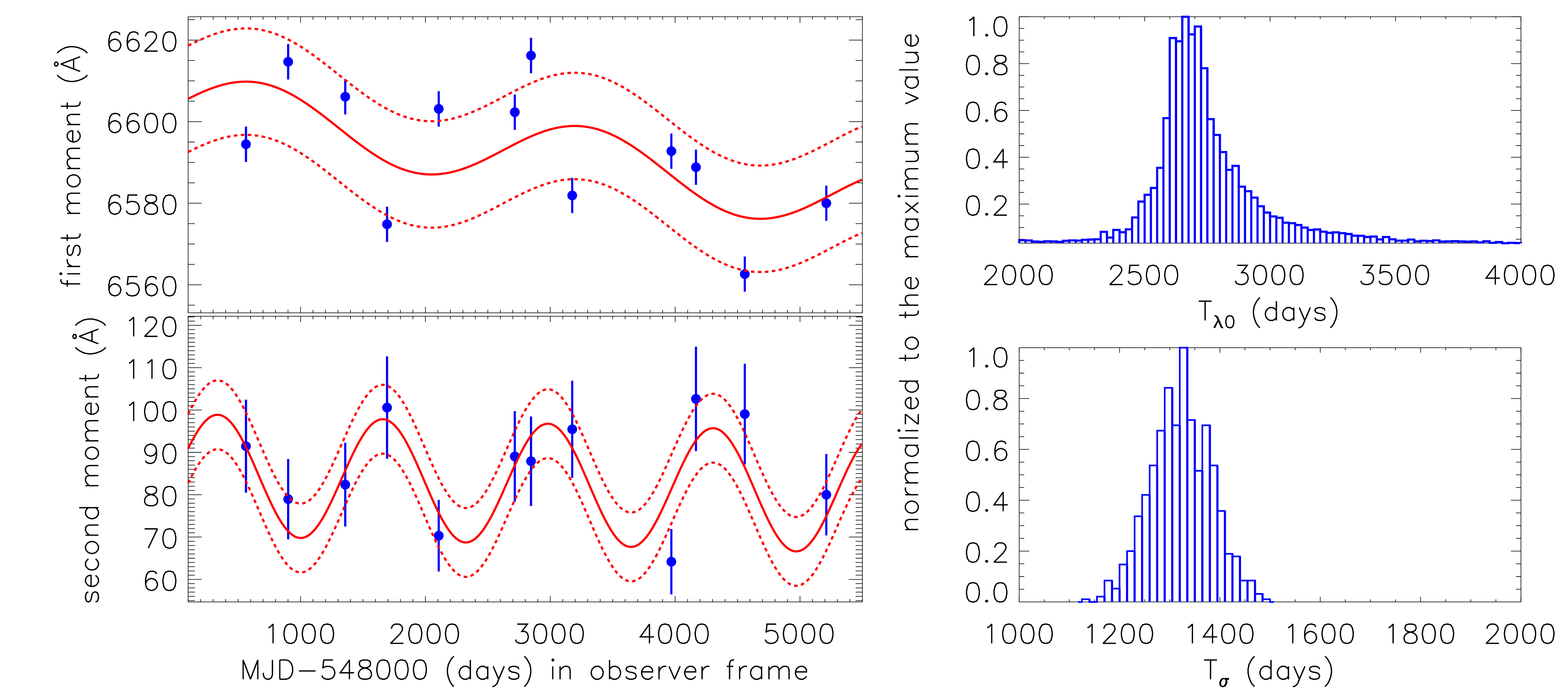}
\caption{Left panels show the variability of the first moment $\lambda_{0}(t)$ and the second moment $\sigma(t)$ of the double-peaked 
broad H$\alpha$ in NGC1097. In the left panels, solid circles plus error bars in blue show the measured results through the 12 high 
quality spectra after host galaxies being removed and narrow emission lines being masked out, solid and dashed red lines show the best 
descriptions and the corresponding 1RMS scatters to the $\lambda_0(t)$ and $\sigma(t)$ determined by the maximum likelihood method 
combined with the MCMC technique. Right panels show the periodicity distributions of $T_{\lambda 0}$ in $\lambda_{0}(t)$ and $T_\sigma$ 
in $\sigma(t)$, respectively.}
\label{n1097}
\end{figure*}

	As noted in the Introduction, both the elliptical accretion disk model and the circular accretion disk plus arms model are 
mainly considered. For the elliptical accretion disk model in \citet{el95}, there are seven free model parameters, the inner radius 
$r_0$ (in units of the Schwarzschild radius $R_G$) and the outer radius $r_1$ (in units of $R_{G}$) of the emission region, the 
inclination angle $i$ of the emission region, the line emissivity power-law index $f_r\propto r^{-q}$, the local turbulent broadening 
parameter $\sigma_L$ (in units of km/s), the eccentricity $e$ of the emission region, and the orientation angle $\omega$. For the 
circular accretion disk plus arms model in \citet{se03}, besides the listed model parameters for the elliptical accretion disk model 
with eccentricity to be zero, there are four additional model parameters, the contrast ratio $A$ for the arms relative to the rest 
of the disk, the width $w$ and pitch angle $p$ for the arms, and the starting radius $r_{s}$ (in units of $R_{G}$) of the arms.

	Then, through the theoretical models, variability of central wavelength (the first moment, $\lambda_0$) and line width 
(the second moment, $\sigma$) can be checked by the following five steps. First, except the model parameter of $\omega$, the other 
model parameters are randomly selected within the accepted limited ranges listed in Table 1. In Table~1, the range from 0.3 to 0.9 
for the $sin(i)$ means the inclination angle larger than 18degrees but smaller than 64degrees, common values for BLAGNs as 
discussed in \citet{zh18}. Second, 30 values of $\omega$ in a cycle are randomly collected from 0 to 2$\pi$. Third, combining 
the 30 values of $\omega$ with the other model parameters, 30 line profiles in one cycle can be created with different $\omega$ 
and with the same other model parameters, within the wavelength from 6100\AA~ to 7200\AA~ for expected double-peaked broad 
H$\alpha$ (6564.61\AA~ as the theoretical central wavelength in rest frame). Fourth, based on the definitions of the first moment 
$\lambda_0$ and the second moment $\sigma$ in \citet{pe04}, the $\lambda_0$ and the $\sigma$ of the 30 simulated line profiles 
can be calculated within one cycle. Fifth, to repeat the four steps above 1000 times, there are 1000 cases with determined 
$\lambda_0$ and $\sigma$ in 30 different phases in one rotation cycle of the disk-like BLRs.

\begin{table}
\caption{Limited ranges for the Model parameters}
\begin{tabular}{llll}
\hline\hline
        par & units & Arms & Elli  \\
\hline\hline
        $r_0$ & $R_G$ & [100,~~600]    & [100,~~600] \\
        $r_1$ & $R_G$ & $r_0$+[300,~~1800] & $r_0$+[300,~~1800] \\
	$\sin(i)$ &   & [0.3, ~~0.9]   & [0.3, ~~0.9] \\
        $q$ &         & [-7, ~~7]  & [-7, ~~7] \\
	$e$ &         & 0 & [0, ~~0.95]   \\
        $\sigma_L$ & km/s & [400, ~~1500]  & [400, ~~1500] \\
        $A$     &     &   [1, ~~3] \\
        $W$     & degree & [10, ~~60] \\
        $p$     & degree & [-45, ~~45] \\
        $r_s$ & $R_G$   & [$r_0$, ~~0.8$r_1$] \\
\hline\hline
\end{tabular}\\
Notice: The first column shows which model parameter is applied. The second column shows the units for the applied model parameters. 
The third column shows the limited range ([lower boundary, upper boundary]) of each model parameter in the circular accretion disk 
plus arms model. The fourth column shows the limited range of each model parameter in the elliptical accretion disk model.
\end{table}

\begin{figure*}
\centering\includegraphics[width = 18cm,height=5cm]{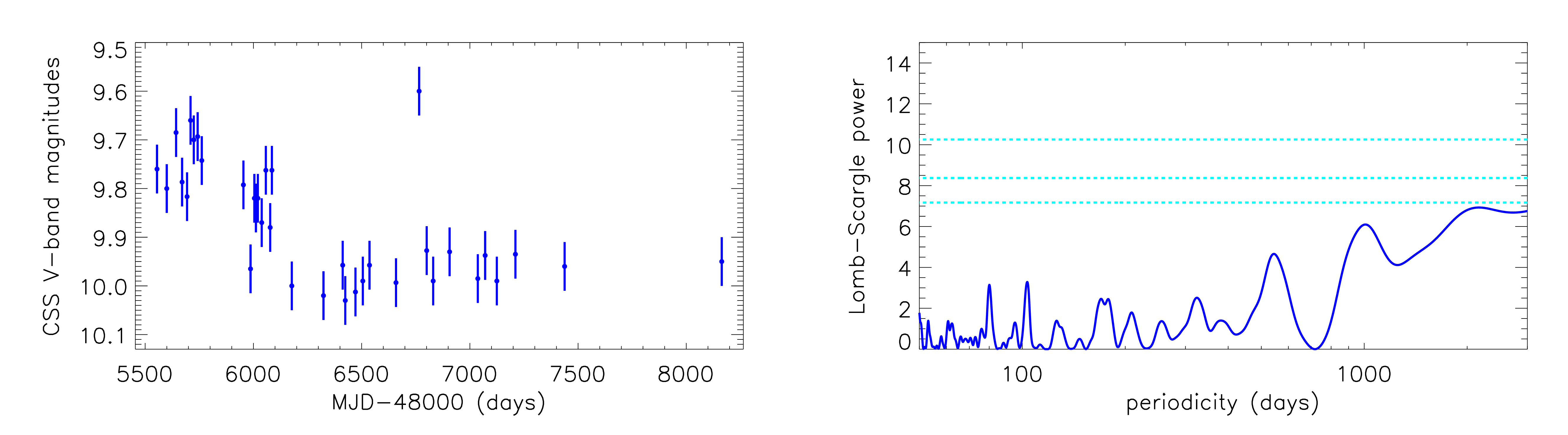}
\caption{Left panel shows the CSDR2 V-band light curve of NGC1097. Right panel shows the corresponding Lomb-Scargle power. In the 
right panel, the horizontal cyan dashed lines from top to bottom show the 90\%, 50\% and 10\% confidence levels (corresponding false 
alarm probability 0.1, 0.5, 0.9), respectively.
}
\label{lmc}
\end{figure*}

	As examples in Fig.~\ref{model}, panel (a) and (d) show simulated lines in different phases in one cycle through the 
elliptical accretion disk model and the circular accretion disk plus arms model. The apparent variability in line profiles can 
be confirmed. Panel (b) and (e) show the corresponding variability of $\lambda_0$ in different phases for the simulated lines 
shown in panel (a) and (d), leading to very clear periodic variations of the $\lambda_0$ in a rotation cycle. Panel (c) and (f) 
show the corresponding variability of $\sigma$ in different phases for the simulated lines shown in panel (a) and (d), leading 
to very clear periodic variations of $\sigma$ in a rotation cycle. Here, rather than a pure sine function, considering $\omega$ 
as the function argument, a sine function plus a linear trend is applied to describe $\lambda_0$ and $\sigma$ on $\omega$ related 
to accretion disk models, as shown in Fig.~\ref{model}.

	In order to check periodic variability of the $\lambda_0$ and $\sigma$ of the simulated double-peaked broad lines, the 
widely accepted Lomb-Scargle periodogram technique \citep{ln76, sj82, zk09, vj18} can be applied to the light curves 
$\lambda_0(\omega)$ and $\sigma(\omega)$ in the 1000 cases through the theoretical accretion disk models. Here, due to no real 
time information for the model created curves of $\lambda_0$ and $\sigma$ on $\omega$, the $\omega$ is applied to trace time 
information. And, through applications of the Lomb-Scargle technique, the searching periodicity range (the range for the x-axis 
in left panels of Fig.~\ref{gls}) is set to be larger than 0.1 (in units of rad) and smaller than 20 (in units of rad), in order 
to show the clear peaks around $2\pi$ (rad). In Fig.~\ref{gls}, panel (a) and (b) show the Lomb-Scargle power of the 
$\lambda_0(\omega)$ and $\sigma(\omega)$ shown in panel (b), (c) and (e), (f) in Fig.~\ref{model}. Interestingly, clear peaks 
with confidence level higher than 99\% (corresponding false alarm probability 0.01) can be found in the Lomb-Scargle power. Then, 
among the 1000 cases through the elliptical accretion disk model and through the circular accretion disk plus arms model, there 
are 860, 894 cases leading to reliable peaks with confidence level higher than 99\% in the Lomb-Scargle power. The corresponding 
peak distributions are shown in panel (c) and (d) in Fig.~\ref{gls}, to support clear periodic variations of the $\lambda_0$ 
and $\sigma$ for the double-peaked broad lines coming from BLRs lying into central accretion disks. Furthermore,  
for all the listed model parameters in the Table~1, the two-sided Kolmogorov-Smirnov statistic technique is applied to confirm 
each model parameter for the 860/894 cases with periodic results and all the 1000 cases having the same distributions with 
probability higher than 99\%. The higher probability more than 86\% (860/1000, 894/1000) for the periodic results through 
simulated results and the same model parameter distributions for the 1000 cases and the 860/894 cases strongly indicate that 
there are few effects of the model parameter space coverage on the detected periodic results.

	Moreover, as the shown results in panel (c) and (d) in Fig.~\ref{gls}, the distributions of $R_{fs}$ (ratio of the 
periodicity in units of rad in $\lambda_0(\omega)$ to that in $\sigma(\omega)$) are shown in panel (e) and (f) in Fig.~\ref{gls}, 
indicating that circular accretion disk plus arms model can lead to $R_{fs}\sim1.98\pm0.08$ (the mean value plus/minus the 
standard deviation), but the elliptical accretion disk model should lead to $R_{fs}\sim1.22\pm0.30$ (the mean value plus/minus 
the standard deviation). The very different $R_{fs}$ can be applied to test which model, elliptical accretion disk model or 
circular accretion disk plus arms model, is preferred to explain the double-peaked broad emission lines.

	Besides the results through the theoretical accretion disk models, it is interesting to check whether the periodic 
variations can be detected in a real double-peaked BLAGN. Here, the known double-peaked BLAGN NGC1097 is collected. Through 
the reported 11 high-quality double-peaked broad H$\alpha$ in \citet{se03} (host galaxy contributions have been removed) and 
the one public spectrum collected from HST (Hubble Space Telescope) mission (ID:8684, PI: Dr. Eracleous) which has been 
described in \citet{zh22a}, after narrow emission lines being masked out, the first moments and the second moments can be 
simply measured in observer frame. The time evolutions of the first moment $\lambda_0(t)$ and the second moment $\sigma(t)$ 
are shown in the left panels of Fig.~\ref{n1097} for the 12 high quality double-peaked broad H$\alpha$ observed from Nov. 2th, 
1991 (MJD=48563) to Jul. 24th, 2004 (MJD=53211).

	Considering a model function including a sine function plus a linear trend applied to describe the $\lambda_0(t)$ and 
the $\sigma(t)$ with the real time $t$ as the function argument, through the maximum likelihood method combining with the Markov 
Chain Monte Carlo technique (MCMC) \citep{fh13}, the best descriptions and the corresponding 1RMS scatters can be determined 
and shown in the left panels of Fig.~\ref{n1097}. The determined robust periodicities are $T_{\lambda0}={2630}_{-55}^{+168}$days 
($T_{\lambda0}$ about $k_{10}\sim47.8$times larger than its negative uncertainty margin 55days and about 
$k_{11}\sim15.6$times larger than its positive uncertainty margin 168days) in $\lambda_0(t)$ and $T_\sigma={1322}_{-64}^{+72}$days 
($T_\sigma$ about $k_{20}\sim20.6$times larger than its negative uncertainty margin 64days and $k_{21}\sim18.4$times 
larger than its positive uncertainty margin 72days) in $\sigma(t)$, as the shown MCMC technique determined posterior 
distributions in the right panels of Fig.~\ref{n1097}, leading the parameter $R_{fs}={1.99}_{-0.14}^{+0.23}$ in NGC1097 well 
consistent with the theoretically simulated results shown in panel (f) in Fig.~\ref{gls}. The results can be applied to support 
that the circular accretion disk plus arms model should be preferred for the double-peaked broad H$\alpha$ in the known 
double-peaked BLAGN NGC1097. Moreover, based on the shown properties of $R_{fs}$ in panel (e) in Fig.~\ref{gls} for elliptical 
accretion disk model, there are only 33 of the 860 cases with $R_{fs}$ larger than 1.85 and smaller than 2.22 (the lower and 
upper values for NGC1097). In other words, the probability is only 3.8\% (33/860) to support the elliptical accretion disk 
model to explain the double-peaked broad H$\alpha$ in the NGC1097. 

	Here, due to only 12 data points in the $\lambda_0(t)$ and $\sigma(t)$ and applications of the sine function plus a 
linear trend leading to the well-accepted descriptions, there are no further discussions on applications of the other 
methods/techniques (such as the Lomb-Scargle technique) to determine periodic results in NGC1097. Furthermore, 
accepted the same time information of $\lambda_0(t)$ and the $\sigma(t)$ shown in Fig.~\ref{n1097}, 100000 artificial couples 
[[$\lambda_0(t,a)$],~[$\sigma(t,a)$]] are created by the corresponding each 12 values of $\lambda_0(t,a)$ and $\sigma(t,a)$ 
randomly collected within the ranges from the minimum $\lambda_0(t)$ to the maximum $\lambda_0(t)$ for each $\lambda_0(t,a)$ 
and from the minimum $\sigma(t)$ to the maximum $\sigma(t)$ for each $\sigma(t,a)$. And, the sine function plus a linear trend 
is applied to determine the probable periodicity $T_{\lambda,a}$ and $T_{\sigma, a}$ in the 100000 artificial 
[[$\lambda_0(t,a)$],~[$\sigma(t,a)$]]. Then, assumed that the determined periodicities in $\lambda_0(t)$ and 
$\sigma(t)$ were not random but intrinsically true in NGC1097, it is necessary to determine how many artificial cases have the 
similar periodic results as those in NGC1097. Based on the following criteria that $T_{\lambda,a}$ and $T_{\sigma, a}$ larger 
than 13 (smaller than $k_{10}$ and $k_{11}$) and 16 (smaller than $k_{20}$ and $k_{21}$) times of 
their corresponding uncertainties and $T_{\lambda,a}$ lying within the range from $2630-55$days to $2630 + 168$days and 
$T_{\sigma, a}$ lying within the range from $1322-64$days to $1322 + 72$days, there are 156 of the 100000 artificial results 
leading to periodicity around 2630days and 1322days in $\lambda_0(t,a)$ and $\sigma(t,a)$, indicating the probability smaller 
than 0.156\% (156/100000) for the detected QPOs being random in NGC1097. Here, not a broad range for periodicities 
in the artificial cases are applied, otherwise, the collected artificial cases should have the ratios of $T_{\lambda,a}$ to 
$T_{\sigma, a}$ very different from the $R_{fs}\sim{1.99}$ in the NGC1097. Therefore, the periodic results in NGC1097 are not 
random but robust enough, at least with confidence level higher than 99.84\% (1-0.156\%).

	Besides the accretion disk origin for the double-peaked broad emission lines, the BBH model \citep{sl10, eb12, de20, 
dc23, sw23} can be simply discussed in NGC1097 as follows. For double-peaked broad lines related to assumed BBH systems, optical 
Quasi-Periodic Oscillations (QPOs) \citep{gd15a, gd15b, zb16, zh24c} could be detected in optical light curves. Unfortunately, 
through the collected 7.1 years-long optical V-band light curve from the Catalina Surveys Data Release 2 (CSDR2) \citep{dd09, 
dd19} with MJD from 53554 (Jul. 2th, 2005) to 56163 (Aug. 23th, 2012) shown in the left panel of Fig.~\ref{lmc}, there are no 
signs for probable QPOs in the CSDR2 light curve, especially through the Lomb-Scargle power shown in the right panel of 
Fig.~\ref{lmc}. The time duration is about 7 years of the CSDR2 light curve of NGC1097, which is comparable to the periodicity 
in $\lambda_0$ in NGC1097. Therefore, the BBH model is not preferred in NGC1097,  due to the none-detected optical QPOs.

	Furthermore, for common accretion disk models, the disk precession period \citep{se03} is around 
$T_p~\sim~1040M_8R_3^{2.5}$years, with $M_8$ as BH mass in units of $10^8{\rm M_\odot}$ and $R_3$ as distance in units of 
$10^3{\rm R_G}$ between BLRs and central BH. Considering the common values \citep{wg24} of $R_3$ not smaller than 0.2, probably 
periodic variations of the first moment and the second moment in around 10years-long multi-epoch optical spectra could be 
expected in double-peaked BLAGNs with central BH masses smaller than $5.4\times{10}^7{\rm M_\odot}$. In the near future, to test 
the periodic variations of the moments of $\lambda_0$ and $\sigma$ should be applied in the double-peaked BLAGN with less 
massive BHs.

	Before ending the section, two points should be noted. First, we have actually tried to collect and check spectroscopic 
results of the other double-peaked BLAGNs in the literature, such as 3C390.3 \citep{zh11} and Arp 102B \citep{sp13, ps14}. However, 
considering the other double-peaked BLAGNs having shorter time spans and/or only several spectra not leading to apparent 
periodic variability in at least one cycle, it is hard to check the periodic results in $\lambda_0(t)$ and in $\sigma(t)$ in 
the other double-peaked BLAGN. Second, as discussed in \citet{bp09, hf20}, observed broad emission lines could include additional 
non-disk component not related to BLRs into central accretion disks, which have important effects on the expected periodic 
results. However, for specific cases, if the non-disk component was from common virialized BLRs, variability of the non-disk 
component should be very smaller than the variability of the disk component, effects of the non-disk component could be few, 
leading to probably similar expected periodic results as those in this manuscript. Unfortunately, at the current stage, there 
is no way to quantify the effects, unless there were clear information of variability intensity ratio of the additional non-disk 
component to the disk component coming from central disk. 

\section{Conclusions}

	The results through theoretical accretion disk models and also through the real observational results in the known 
double-peaked BLAGN NGC1097 strongly indicate that the periodic variations of the first moment and the second moment of the 
broad emission lines can be accepted as signs for the broad emission lines coming from BLRs lying into central accretion disks 
in BLAGN. Moreover, the periodicity ratio around 2 from the periodic variations of the first moment to the periodic variations 
of the second moment can be accepted as the signs to support the structures of spiral arms in the BLRs in the double-peaked 
BLAGNs. The results further provide an independent method to test the accretion disk origin of the double-peaked broad emission 
lines only through the moments from the profiles of broad emission lines, without considering physical properties of theoretical 
model determined parameters.

\begin{acknowledgements}
Zhang gratefully acknowledge the anonymous referee for giving us constructive comments and suggestions to greatly 
improve the paper. Zhang gratefully thanks the kind financial support from GuangXi University and the kind grant support from 
NSFC-12173020 and NSFC-12373014. This manuscript has made use of the NASA/IPAC Extragalactic Database (NED) operated by the Jet 
Propulsion Laboratory, California Institute of Technology, under contract with the National Aeronautics and Space Administration.
\end{acknowledgements}

\end{document}